\documentclass{elsarticle}
\makeatletter
\def\ps@pprintTitle{%
 \let\@oddhead\@empty
 \let\@evenhead\@empty
 \def\@oddfoot{\centerline{\thepage}}%
 \let\@evenfoot\@oddfoot}
\makeatother

\usepackage{lineno}

\usepackage[table]{xcolor}
\usepackage{graphicx}
\usepackage{rotating}
\usepackage{pict2e}

\usepackage{dcolumn}
\usepackage{bm}
\usepackage[hidelinks]{hyperref} 
\usepackage[all]{hypcap}

\usepackage{relsize}

\usepackage{booktabs}

\usepackage{mathtools}

\newcommand{\enquote}[1]{``#1''} 

\usepackage{xspace}

\newcommand{\ev}{{\rm e}\kern-1.pt{\rm V}}
\newcommand{\gev}{{\rm Ge}\kern-1.pt{\rm V}}
\newcommand{\mev}{{\rm Me}\kern-1.pt{\rm V}}
\newcommand{\kev}{{\rm ke}\kern-1.pt{\rm V}}
\newcommand{\tev}{{\rm Te}\kern-1.pt{\rm V}}
\newcommand{\gevsq}{\mbox{$\mathrm{{\rm Ge}\kern-1.pt{\rm V}}^2$}}

\def\lsim{\mathrel{\rlap{\lower4pt\hbox{\hskip1pt$\sim$}}
    \raise2pt\hbox{$<$}}} 
\def\gsim{\mathrel{\rlap{\lower4pt\hbox{\hskip1pt$\sim$}}
    \raise2pt\hbox{$>$}}} 

\begin{document}


\title
{
  Quantitative Assessment of Finite-element Models for Magnetostatic Field Calculations
}

\author[Cornell]{J.A.~Crittenden\corref{cor1}}
\ead{crittenden@cornell.edu}
\address[Cornell]{CLASSE\fnref{fn1}, Cornell University, Ithaca, NY 14853, United States}
\cortext[cor1]{Corresponding author. Tel.: +1 6072554882}

\date{\today} 

\begin{abstract}
  We present quantitative means for assessing the numerical accuracy 
  of static magnetic field calculations in finite-element models. Our calculations
  use the three-dimensional Opera simulation software suite of
  Dassault Syst\`emes. Our  need to assess the effects of fringe fields
  requires such a 3D algorithm. While we do discuss and compare our approach to a method of
  accuracy estimation used in the Opera post-processor, our methods are generally
  applicable to any model
  using relaxation techniques in finite-element systems. For purposes of
  illustration, we present modeling and analysis of two types of quadrupole electromagnets
  presently in operation in the south arc of the Cornell Electron Storage Ring (CESR).
  Calculations of field multipole expansion coefficients and numerical deviations from Maxwell's equations
  in source-free regions are discussed, with emphasis on the dependence of their accuracy
  on changes to the finite-element model.
  Successive refinement steps in the finite-element model for the non-extraction type
  of CESR south arc quadrupole achieve a reduction in the RMS value of the
  longitudinal component of the curl vector on the magnet axis by a factor of nearly~70 from
  $4.38 \times 10^{-2}$~T/m to $6.36 \times 10^{-4}$~T/m, which is 0.0021\% of the field gradient.
   An accuracy of 2.9\% is achieved for a dodecapole
   coefficient of $2.4 \times 10^{-4}$ of that of the quadrupole in the longitudinal integral of
   the vertical field gradient at a radius of 1~cm.
  
\end{abstract}

\maketitle

\clearpage

\tableofcontents


\section{Introduction}

\begin{figure}[htb]
      \centering
        \includegraphics[width=0.35\columnwidth]{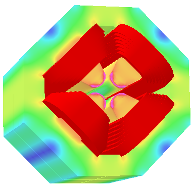}
        \includegraphics[width=0.20\columnwidth]{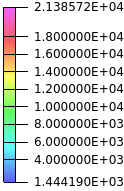}
        \includegraphics[width=0.35\columnwidth]{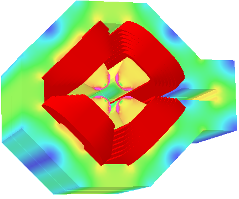}
        \caption{Color maps of the magnetization field vector magnitudes calculated in the Opera model on the non-extraction (Type~A, left)
          and extraction (Type~B, right) south arc quadrupole magnets at nominal field intensity. The units are Gauss. Some effects of saturation are evident
          in the ends of the pole tip.
        }
        \label{fig:colormaps}
    \end{figure}

Accurate modeling of static magnetic fields plays an important role in the design, construction,
commissioning and operations for particle accelerators. We have developed means
for quantitative assessment of the accuracy of field calculations in
finite-element-based models. General treatments of the finite-element method and engineering applications
can be found, for example, in Refs.\cite{strang2,zienkiewicz,akin}. 
Here we use the examples of two types of quadrupole magnet operating in
the CHESS-U upgrade of the Cornell Electron Storage Ring
(CESR)\cite{PhysRevAccelBeams.22.021602}.
The two quadrupoles at the entrance and exit of each of six
achromat sections in the new south arc are designed for exact quadrupole symmetry (Type~A),
while the two central quadrupoles are equipped with extraction channels
on the outside of the ring (Type~B). 
Figure~\ref{fig:colormaps}
shows color maps of the magnetization field vector magnitude on the steel
surfaces of the Opera models described in this report.

One motivation for this work is to assess the accuracy of the idealized approximate
models of our quadrupole magnets in the
development and operations lattice model based on the Bmad library\cite{Sagan:Bmad2006}, and to determine quantitatively its adequacy for any given purpose. These models consist merely of values for the field gradient and length, chosen such that their product is consistent with modeling and field measurements. The assessment can be done by comparison either with tracking through modeled field maps (see Refs.\cite{Meot:2019lor,Meot:ICAP2018-TUPAF08} for a recent example of lattice analysis using field-map tracking) or by adding field integral multipole terms to the Bmad lattice element.
The latter method is orders of magnitude faster in computation speed, but its accuracy must
first be verified by comparison with reliable tracking through field maps. Thus knowledge of the numerical accuracy of multipole calculations in the model is required. Our approach is described in Sec.~\ref{sec:diagnostics:multipoles}.
We make use of the symmetry constraints for the quadrupole-symmetric magnet Type~A to obtain estimates for the numerical uncertainties inherent in the refined finite-element model. These can then be used to estimate the accuracy in the
multipole coefficients for the quadrupole magnet Type~B. By comparing multipole expansions in the vertical field
component at the center of the magnet to that of the integrated field component, we can distinguish
multipole contributions by the pole shape from those contributed by the fringe field.

Our second criterion for estimating the accuracy of the models is an analysis of the numerical violations of Maxwell's equations in a volume relevant to the particle trajectories, as described in Sec.~\ref{sec:diagnostics:maxwellviolations}. The divergence and the three components of the curl of the electric field vector deviate from zero primarily in the fringe field region in our models. We calculate the RMS value of each of these quantities along longitudinal lines
as a function of horizontal position to estimate the improvement in magnitude and location of the errors as the
finite-element model is refined. Since the longitudinal sums of the errors tend to be small, we use the RMS values,
since all errors contribute to errors in the modeled particle trajectory. We employ a difference-over-sum method to obtain
relative deviations along the longitudinal coordinate, then propagate those relative errors to obtain an uncertainty in the
RMS value as a function of transverse position. This dependence of non-zero divergence and curl values on the horizontal coordinate motivates various types of changes to the
finite-element model.

These diagnostic criteria are applied to our Opera models for the south arc
quadrupoles in Sec.~\ref{sec:feamodel}, showing their evolution
at each step of the progressive refinement of the finite-element algorithms used  in the volumes and surfaces of the steel and air structures.


\section{Properties of the Quadrupole magnets in the CESR South Arc}
The CHESS-U south arc optics include 24 horizontally focusing quadrupole electromagnets.
These magnets were designed, built and installed  by  S.~Barrett, A.~Lyndaker and A.~Temnykh in collaboration with industrial partners and the CLASSE technical staff.
The dimensions and nominal magnetic field and electrical parameters of these magnets are shown in Table~\ref{tab:quadparams}.
\begin{table*}[hbt]
\centering
  \small
  \label{tab:quadparams}
\caption{Dimensions and nominal operating parameters used in the Opera models for the CESR south arc quadrupole magnets}
\linethickness{3mm}
\tabcolsep 2mm
\begin{tabular}{|l|c|c|}
\hline
{\bf {\small Parameter}} &  {\bf {\small Type A}} & {\bf {\small Type B}}\\
&{\bf {\small non-extraction}} & {\bf {\small extraction}}\\
\hline
Number of magnets & 12 & 12\\
\hline
Bore radius (cm) & \multicolumn{2}{c|}{2.30}\\
\hline
Steel height (cm) & \multicolumn{2}{c|}{47.44}\\
\hline
Steel width (cm) & 47.44 & 56.72 \\
\hline
Steel length (cm) & 37.70 & 33.30\\
\hline
Length including coil (cm) & 54.33 & 49.94\\
\hline
Pole width (cm) & \multicolumn{2}{c|}{8.56}\\
\hline
Central Field Gradient (T/m) & -30.36 &  -28.01 \\
\hline
Field Gradient Integral (T) & -12.31 & -10.13\\
\hline
Good Field Region$^*$ (mm) & \multicolumn{2}{c|}{$\pm$~10}\\
\hline
Field Gradient Uniformity (\%) & \multicolumn{2}{c|}{0 - 0.1}\\
\hline
NI per coil (Amp-turns) & 6579 &  6033 \\
\hline
Turns per coil & \multicolumn{2}{c|}{($4 \times 8) + 7 + 6 + 5 + 4 + 3 + 2 = 59$}\\
\hline
Conductor cross section (cm x cm) &  \multicolumn{2}{c|}{$0.635 \times 0.635$ with 0.318 diameter hole} \\
\hline
Conductor straight length (cm) & 37.7 - 47.28 & 33.27 - 42.78\\
\hline
Coil inner corner radius (cm) & \multicolumn{2}{c|}{1.96}\\
\hline
Conductor length per turn, avg (cm) & 34.6 & 30.6\\
\hline
R$_{\rm coil}$~ ($\Omega$) & 0.0214 & 0.0198 \\
\hline
L (mH) & 4~x~27  & 4~x~24 \\
\hline
Power supply current (A) & 111.5 &  102.3\\
\hline
Current density (A/mm$^2$) & 2.02 &  1.85 \\
\hline
Voltage drop/magnet (V) & 9.54 & 8.10\\
\hline
Power/magnet (W) & 1064 &  829 \\
\hline
\multicolumn{3}{l}{$^*$ Defined as relative deviation from the central field gradient integral less than $0.1\%$.}
\end{tabular}
\end{table*}
\clearpage
\section{The Opera Application Suite}
\label{sec:opera}
\subsection{Modeler}
\label{sec:opera:modeller}
The construction of an Opera model for an electromagnet can be categorized
in three parts:
\begin{itemize}
\item
  Steel volumes. These volumes can be reduced according to the symmetry of the model. For example, in the case of the Type A quadrupole, a 1/16 model is possible, and the symmetries about the XY, YZ, ZX and 45-degree planes\footnote{The horizontal, vertical and longitudinal coordinates are denoted X, Y and Z, respectively. The origins are at the center of the quadrupole geometry.} in the field map are guaranteed, since field values are calculated only in the region of the 1/16 model.



  Various native geometrical shapes are provided with optimized meshing algorithms. Algebraic functions can be modeled as well, using the MORPH command. The latter can be used for quadrupole pole face shapes, as can a list of 3D points, using the WIREEDGE command.  The Modeler can import files in Spatial's ACIS solid modeling format (.SAT) files) but the meshing can be much less efficient and much more time consuming. Indeed, if the SAT file is too complicated, e.g. with many very small surfaces, the mesh procedure can fail entirely. In our present case, simplifying modifications to an SAT file\footnote{Made available by A.~Lyandaker, CLASSE} were required to
  achieve reliable meshing,\footnote{We acknowledge valuable assistance from Y.~Zhilichev, Dassault Syst\`emes.} i.e. meshing success robust against the variety of modifications to the model described in Sec.~\ref{sec:feamodel}.
  
  The accuracy of the finite-element model and overall field calculation is variously sensitive to the parameters of maximum cell size and cell geometry type defined for each volume. These must be chosen judiciously in the interest of model size and computation time as well.

  The magnetic properties of the steel material can be provided in the form of a table of values relating magnetic flux density values to magnetic field values. The south arc quadrupoles were made of a steel similar to the low-carbon~1010 steel.
\item
  Air volumes. The encompassing volume must be chosen large enough that the
  (automatic) outer tangential magnetic boundary conditions have negligible effect on the field calculations in the regions of interest. Air volumes in the regions of interest, such as those relevant to the particle trajectories,
  must be chosen to be of sufficient refinement for the desired accuracy. Since these two types of volume have large differences in cell size and volume, buffer volumes of intermediate cell size may be required to aid in the meshing convergence.

  The comments above on maximum cell size and type in the list item concerning steel volumes apply to the air volumes as well. It is to be noted, however, that the Modeler meshing algorithm intrinsically adjusts its result to the local derivatives in the steel volume. As a consequence, reducing the maximum cell size parameter does not have the third-power effect on model size and computing time that it does in the air volumes. We will also see in Sec.~\ref{sec:feamodel} that adjusting cell sizes on surfaces can be effective in improving model accuracy without impractical increases in model size and computation time.
\item
  Conductors. The model is required to contain the full 3D geometry of the conductors, even when the steel volumes take advantage of available symmetries.
  In the case of coils arranged with symmetry about the Z axis, provision is made for defining the coil geometry once and specifying the symmetry.
  A wide variety of native geometries are available. In the case of our quadrupoles, we
  modeled each of the four coils as seven racetracks, blocks of 8x4, 7x1, 6,1, 5x1, 4x1, 3x1, and 2x1 1/4-inch-square conductors giving the 59 turns in trapezoidal shape of the coil.
\end{itemize}

\subsection{Solver}
\label{sec:opera:solver}
Magnetostatic calculations are implemented in the Opera package via the TOSCA program which originated
at the Rutherford Laboratory in the 1960's. Its development and many applications are documented in the proceedings
of the Compumag conferences, which are organized by the non-profit International Compumag Society.
The physical basis for the TOSCA algorithm is described by Simkin and Trowbridge in their
contribution to the 1976 Compumag proceedings\cite{compumag1976:simkin}.

Invocation of the default iterative TOSCA program is largely parameter-free,
the memory use and computing time determined by the
size of the finite-element model defined in the Modeler. The calculation has multi-core capability.
There is an option to use a direct solver rather than the iterative method, with attendant greater memory requirements.
The default results of the TOSCA calculation are values for the magnetic scalar potential, magnetic field vector and magnetic flux density vector at each node in the model. There is provision for running multiple field calculations with varying excitation current for the purpose of producing saturation curves.

\subsection{Post-processor}
\label{sec:opera:postprocessor}
The Post-processor can calculate and write tables on a regular grid on the basis of the TOSCA result in two ways. The first way is to interpolate the field value at any point by averaging over nearby nodes. The second method is to integrate over all current density and steel magnetization sources. In our case of calculating field values in the beam region, this latter method produces, by construction, small local derivatives in the result, since the sources are far away
and closer to each other compared to the table step size. Such a result can appear attractive, since the residuals of a polynomial fit for multipole coefficients, for example, are smaller than those from the local node interpolation method.
However, the overall accuracy of the field calculation depends on the mesh accuracy {\em everywhere}. A finer mesh at the beam improves the field calculation at the distant nodes as well. So, perhaps counter-intuitively, refinement the local mesh at the beam can improve the accuracy of the integration method. The integration method, despite the smoothness of its result, requires a sufficiently detailed mesh at the beam, which obviously improves the method of nodal interpolation. Thus the use of the integration method does not by itself improve the accuracy of the
field values in the table. Mesh tuning is necessary in either case, and since the nodal interpolation method must reach the required accuracy in any case, the integration method can be considered superfluous. Indeed, the original motivation for including the integration method option was not for improved local accuracy, but rather to provide better field calculations in volumes far from the steel, where the mesh may be very coarse.\footnote{Private communication from Y.~Zhilichev, Dassault Syst\`emes}

The integration method can be misleading in another way as well. When boundary conditions are defined on symmetry planes in the beam region, the integration method provides an exact constraint by definition, because there are no sources
on the constrained plane. For example, the vertical field component on the magnet axis from magnetization sources is calculated to be zero by definition, since the sources are only available in the 1/16 model, far from the beam. This provides a distortion of the numerical model on the plane, with nodes neighboring the plane exhibiting larger derivatives to the plane than to other neighboring nodes. Again, the only way to improve obedience to the symmetry condition is to refine the mesh. The method of nodal interpolation produces violations of Maxwell's equations on the symmetry planes consistent with the accuracy of the finite-element model.

So what criteria can be used to assess the accuracy of the field calculations
in the field map? One idea might be to use the difference between the two types of field calculations, but the above discussion shows that such a comparison is fraught with systematics unrelated to the overall accuracy of the model.
The Opera Post-processor provides a system variable which addresses the local accuracy of the field calculation. It compares the nodally interpolated field value to an average of scalar potential derivatives in the nodes making up the local cell. This variable, however, provides only a value for the numerical uncertainty in the magnitude of the magnetic field vector. Its intended purpose is to reveal regions in which the mesh requires refinement by comparison to other regions, not to provide a quantitative estimate of the field calculation accuracy. The quest for such a quantitative estimate motivated the present work. Our approach is described in Sec.~\ref{sec:diagnostics} below.

\section{Field Calculation Diagnostic Criteria}
\label{sec:diagnostics}
\subsection{Phantom Multipoles}
\label{sec:diagnostics:multipoles}
Numerical errors in our finite-element model can result in a multipole fit
giving disallowed multiple coefficients (``phantom multipoles'') even in a perfectly symmetric
geometry, as in the case of the south arc quadrupole Type~A. As discussed in
Sec.~\ref{sec:opera:postprocessor}, the integration method of field calculation
can enforce the symmetry at the cost of anomalous discontinuities, but for our application we wish to use the violations of symmetry in our field map as a measure of the numerical accuracy of the
successive models as we refine them (see Sec.~\ref{sec:feamodel}), so we use the nodal interpolation method of calculating the field values.
Confusion can also arise from correlations between the coefficient values in the polynomial fit, so it is advantageous to omit poorly determined coefficients. For the purposes of this illustration, we omit the skew multipole coefficients and fit a polynomial to the horizontal dependence of the longitudinal integral of the vertical field component.
The field map extends over a horizontal region of -10~mm$<$X$<$10~mm in 21 1-mm-wide steps and over a
longitudinal region -300~mm$<$Z$<$300~mm in 301 2-mm-wide steps.\footnote{The granularity of the field map is an additional source of numerical errors, beyond those presented by the finite-element model. While we do not address these in this work, we note that the diagnostic methods described here can be used to quantify their contributions as well.}

The result of a number of trial procedures is the following:
\begin{enumerate}
\item
  Perform a polynomial fit to $\int B_{\rm Y}({\rm X},{\rm Y}=0,{\rm Z})\;{\rm dZ}$. The result of this
  step for the final model in our refinement procedure is shown in Fig.~\ref{fig:fitprocedure}~a).
      \begin{figure}[htb]
        \centering

        \includegraphics[width=0.42\columnwidth]{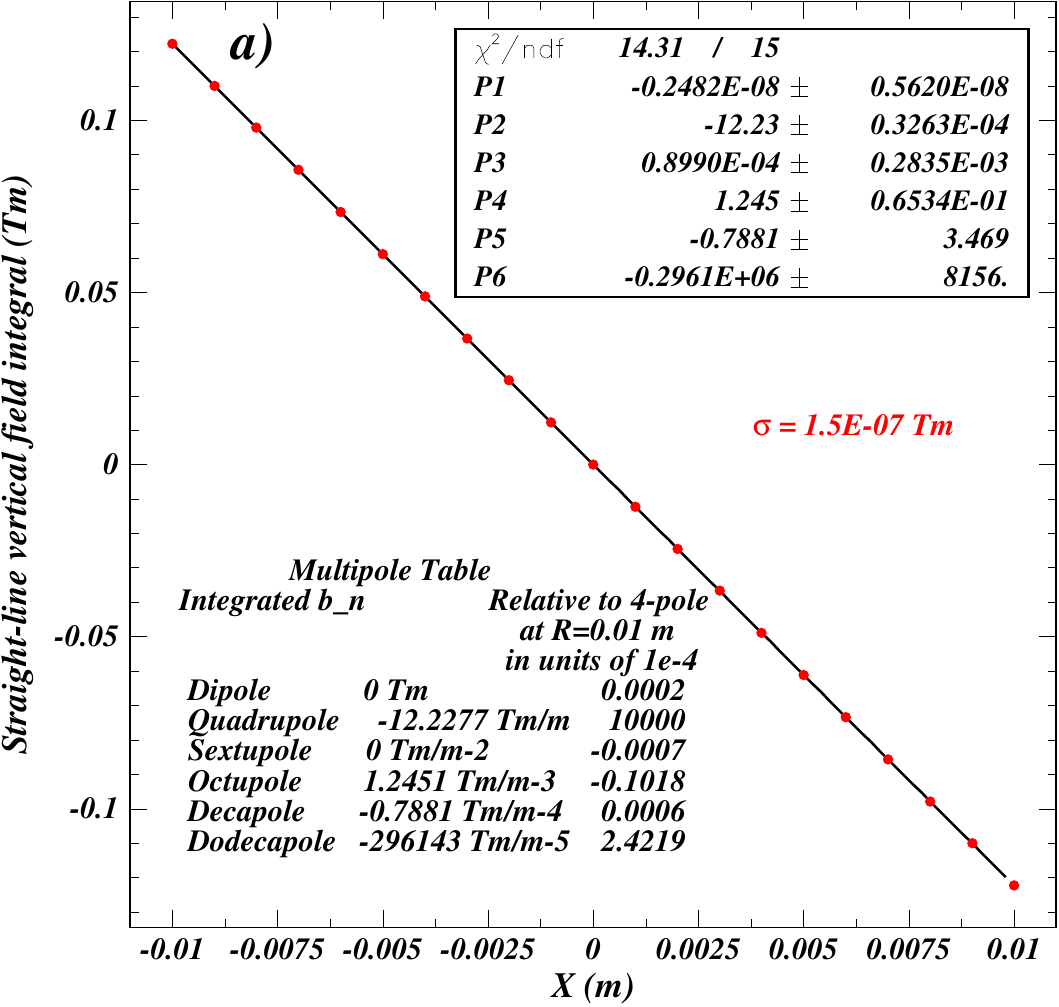}
                    \includegraphics[width=0.42\columnwidth]{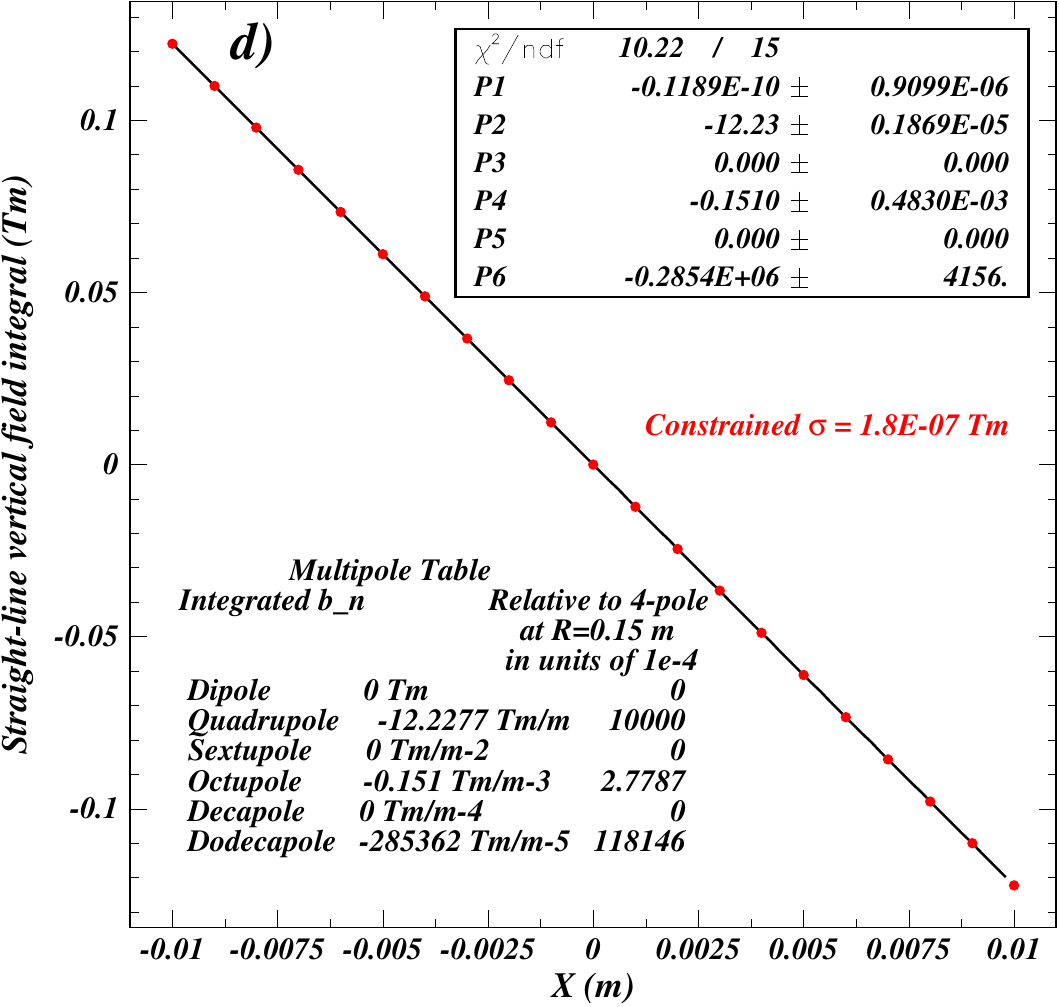}\\

              \includegraphics[width=0.42\columnwidth]{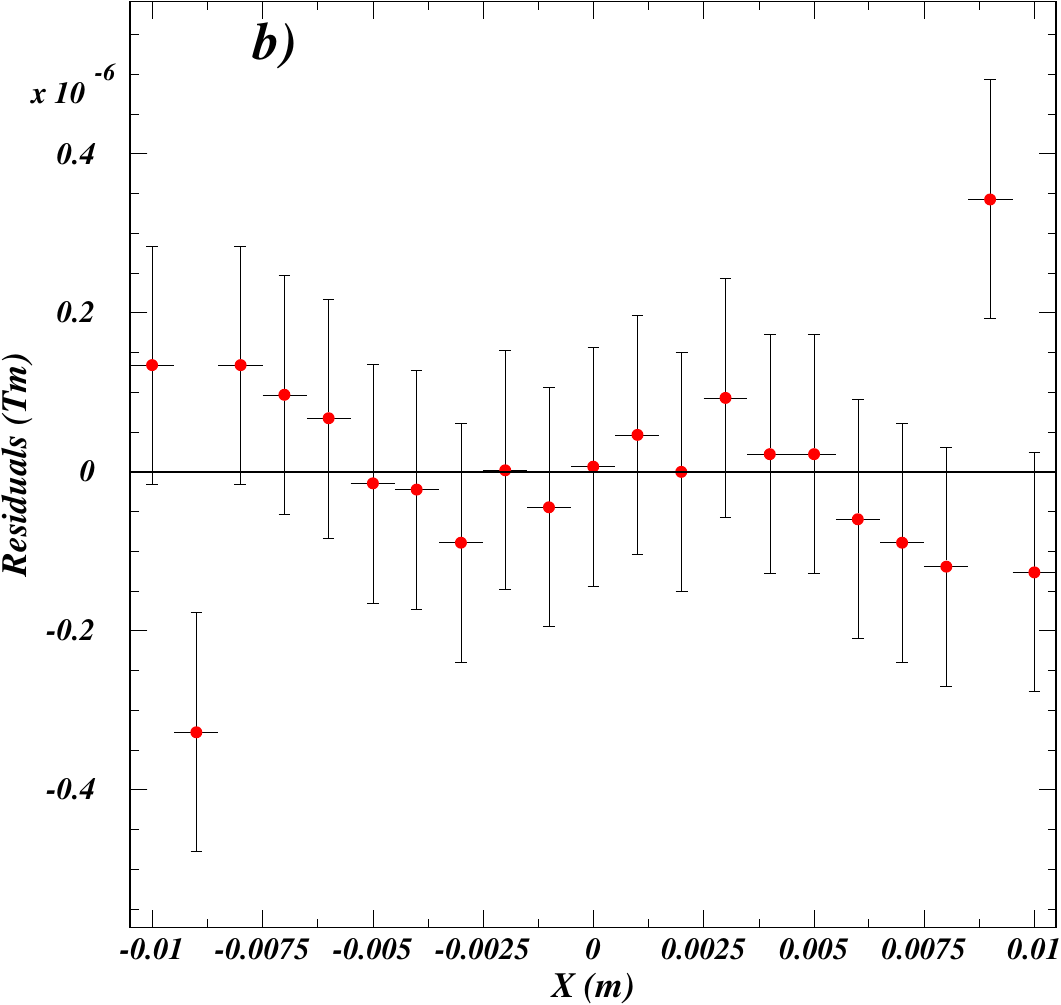}
             \includegraphics[width=0.42\columnwidth]{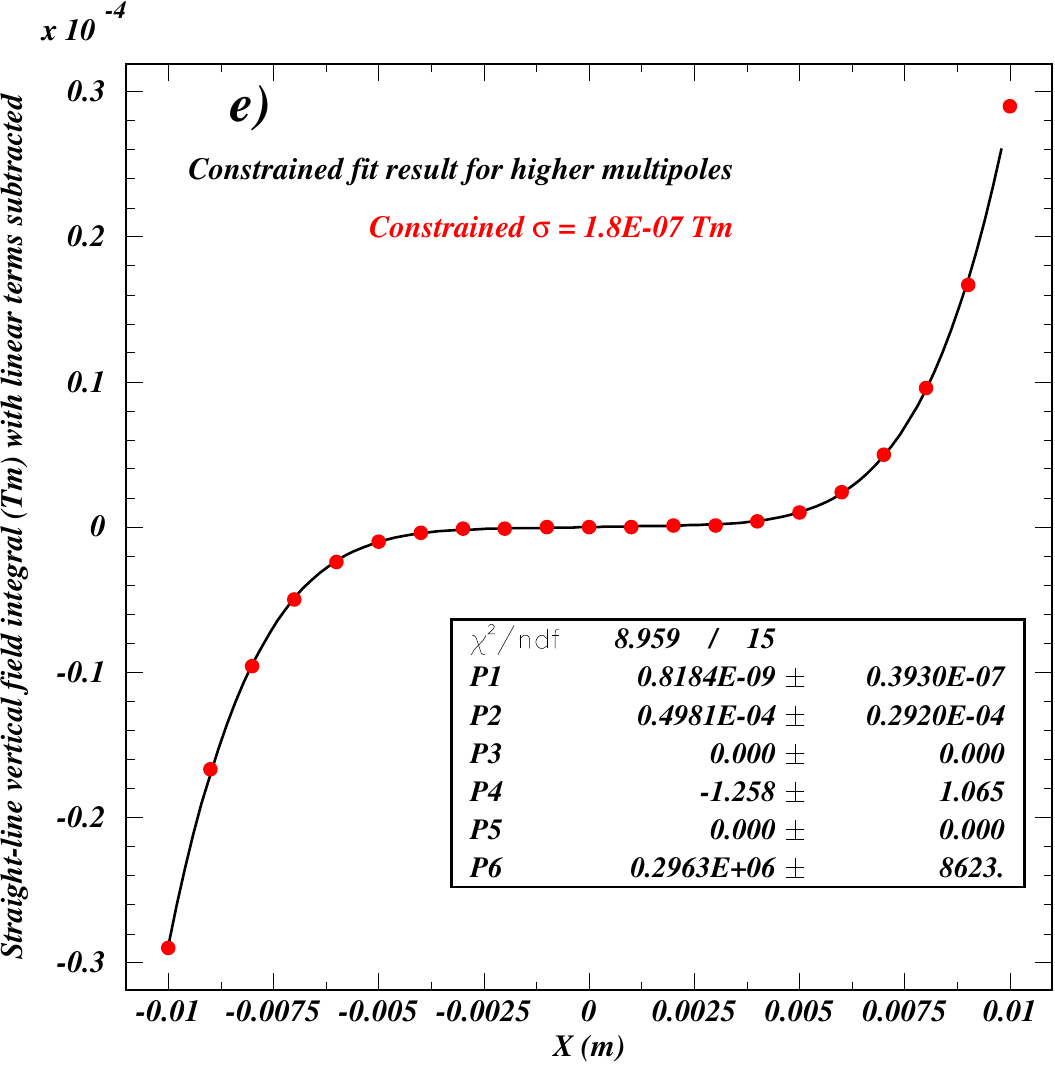}\\

                    \includegraphics[width=0.42\columnwidth]{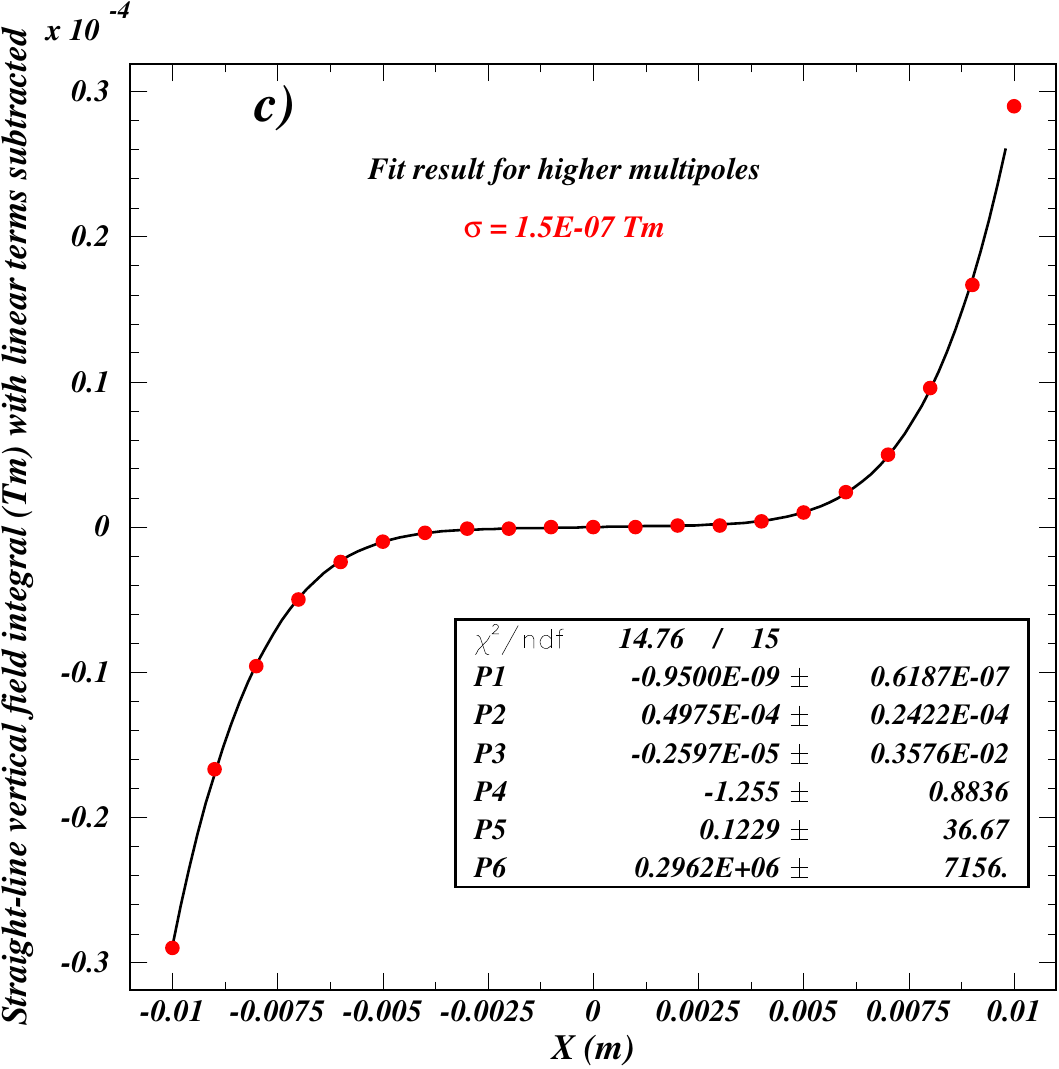}
               \includegraphics[width=0.42\columnwidth]{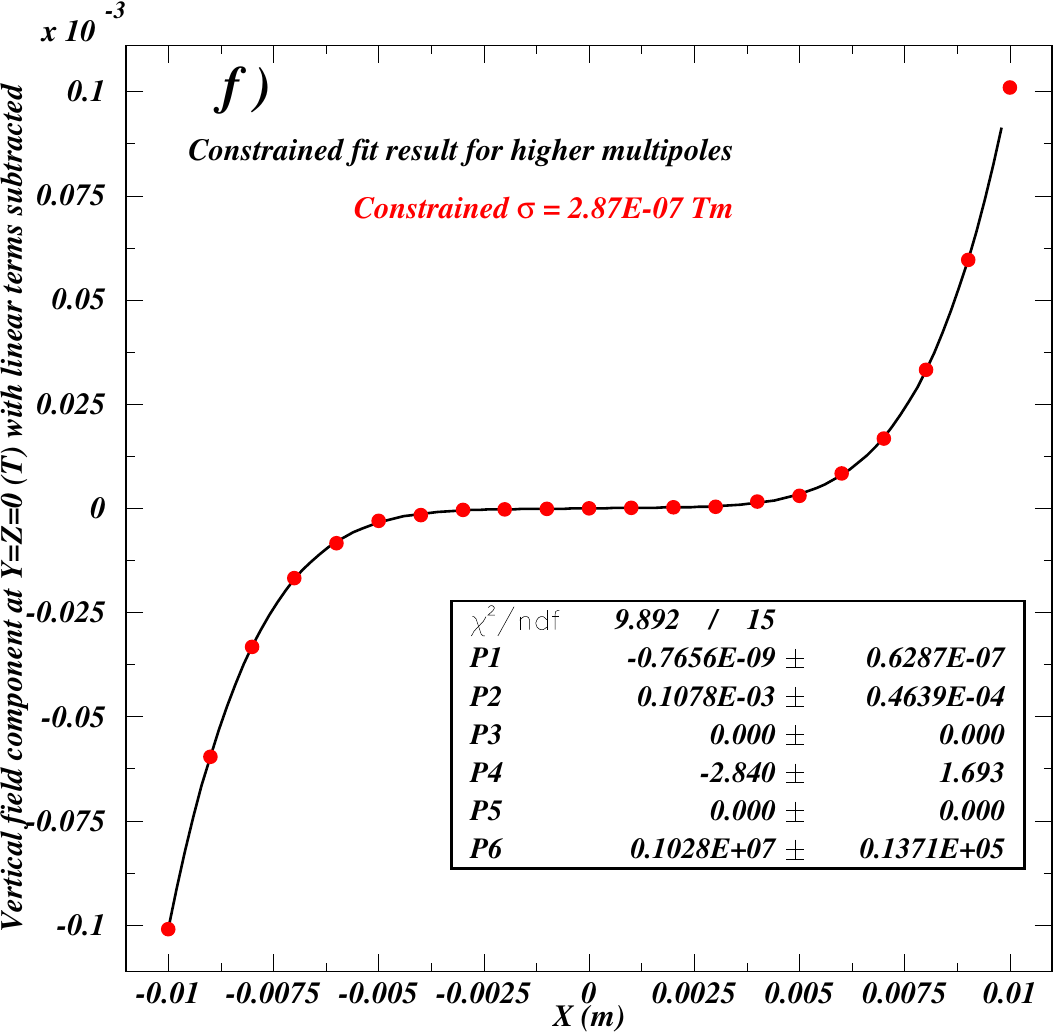}

              \caption{a)~Initial fit for the multipole coefficients in $\int B_{\rm Y}({\rm X},{\rm Y}=0,{\rm Z})\;{\rm dZ}$. The weights in the $\chi^2$ calculation are chosen to give $\chi^2$/NDF near unity so that the uncertainties given for the coefficient calculations are reasonable. b)~Residual distribution resulting from the fit result shown in Fig.~\ref{fig:fitprocedure}~a). The lack of a smooth power-law structure indicates that no terms need to be added to the polynomial. c)~Fit result for the higher multipoles with the linear contributions subtracted. d)~Polynomial fit to the full distribution with the insignificant (or poorly determined) multipole terms constrained to zero. e)~Results of the fit for higher multipoles after the linear part has been subtracted and the insignificant multipole terms have been forced to zero. f)~Result of the procedure to determine the higher multipole content, now applied to the central field $B_{\rm Y}({\rm X},{\rm Y}=0,{\rm Z})$ rather than to its longitudinal integral. The larger dodecapole terms shows that the fringe field compensates the term in the central field, reducing the term in the integral by a factor of nearly three.}
        \label{fig:fitprocedure}
    \end{figure}
      We use the optimization package MINUIT\cite{JAMES1975343,minuit} with the enhanced uncertainty evaluation referred to as MIGRAD+HESSE+MINOS\cite{mnerror,BRUN1989432,paw} for determining the accuracy in the polynomial coefficient determinations.The MINUIT-reported parameter uncertainties are defined as the change in the parameter which yields a change of unity in $\chi^2$/NDF. After convergence, the full second-derivative matrix of the $\chi^2$ function is calculated using a finite-difference method and inverted. The effects of correlations between parameters are therefore included in the calculation. Multiparameter uncertainties are discussed in more detail in Ref.\cite{mnerror}. In the procedure used here, the single value chosen for the weight on each squared residual in the $\chi^2$ calculation is first arbitrarily set to 1~(Tm)$^{-2}$, then the fit is repeated with weight set to a value such that $\chi^2$/NDF is close to unity, as shown in Fig.~\ref{fig:fitprocedure}~a). The corresponding weight value
in this case is $\sigma^{-2}=(1.5 \times 10^{-7}$~Tm)$^{-2}$. We monitor this measure of the scatter of the field integral around the fit result during the model refinement procedure described below in Sec.~\ref{sec:feamodel}.
\item
  Now consider the distribution of the residuals as shown in Fig.~\ref{fig:fitprocedure}~b).
    If any smooth periodic structure indicates the need for additional terms in the polynomial, repeat the fit with more terms. Use as few terms as possible to avoid complicated correlations between the results for the coefficients.
\clearpage
  \item
    Subtract the result for the linear terms from the polynomial and fit again, allowing all coefficients to vary. This provides a more accurate determination of the higher multipoles. Figure~\ref{fig:fitprocedure}~c) shows the result in the present example.
    Select any coefficients smaller than the uncertainty in their determination and force them to zero. In the present example, we have a sextupole term of
    \mbox{$-2.6 \times 10^{-6} \pm 3.6 \times 10^{-3}\;\rm{Tm/m}^2$} and a decapole term of  \mbox{$0.1  \pm 36.7\;\rm{Tm/m}^5$}.
  \item
    Repeat the full fit with these coefficients constrained to zero and reset the weight
    value to obtain $\chi^2$/NDF near unity. This is a small correction. The result is shown in Fig.~\ref{fig:fitprocedure}~d).
  \item
Finally, subtract the linear terms and fit to find the definitive values for the higher-order multipoles, as shown in Fig.~\ref{fig:fitprocedure}~e). We find significant octupole and dodecapole coefficients
of \mbox{$-1.26 \pm 1.07\;\rm{Tm/m}^3$} and \mbox{$2.963 \pm 0.086 \times 10^{5}\;\rm{Tm/m}^5$.}
\end{enumerate}

The contribution from the fringe field can be identified by applying the
above procedure to the central field component
$B_{\rm Y}({\rm X},{\rm Y}=0,{\rm Z})$
The result, shown in Fig.~\ref{fig:fitprocedure}~f),
indicates that the dodecapole term is primarily sourced by the internal pole face shape, with the fringe field reducing its contribution to the integral by a factor of about three. The octupole term is relatively poorly determined, preventing any conclusion.

\begin{figure}[htb]
      \centering
      \includegraphics[width=0.45\columnwidth]{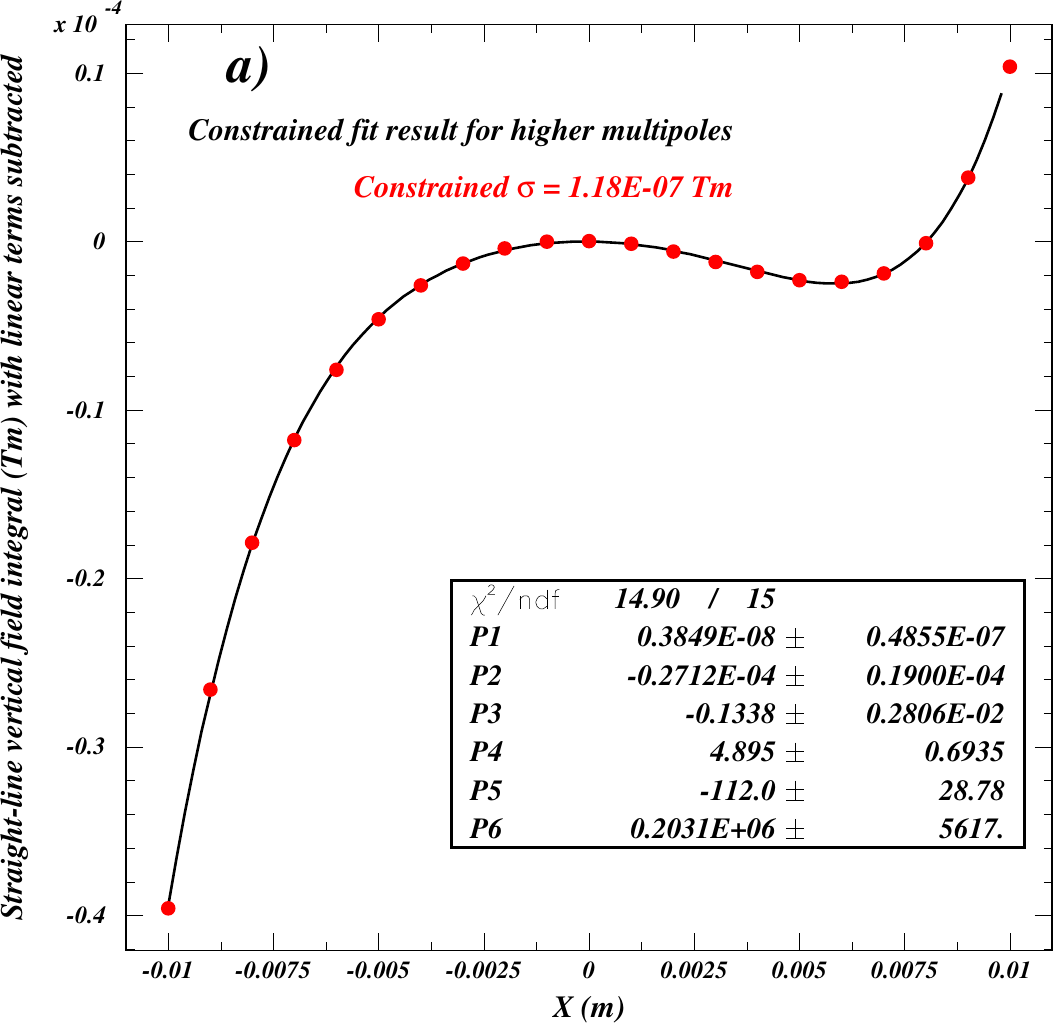}
      \includegraphics[width=0.45\columnwidth]{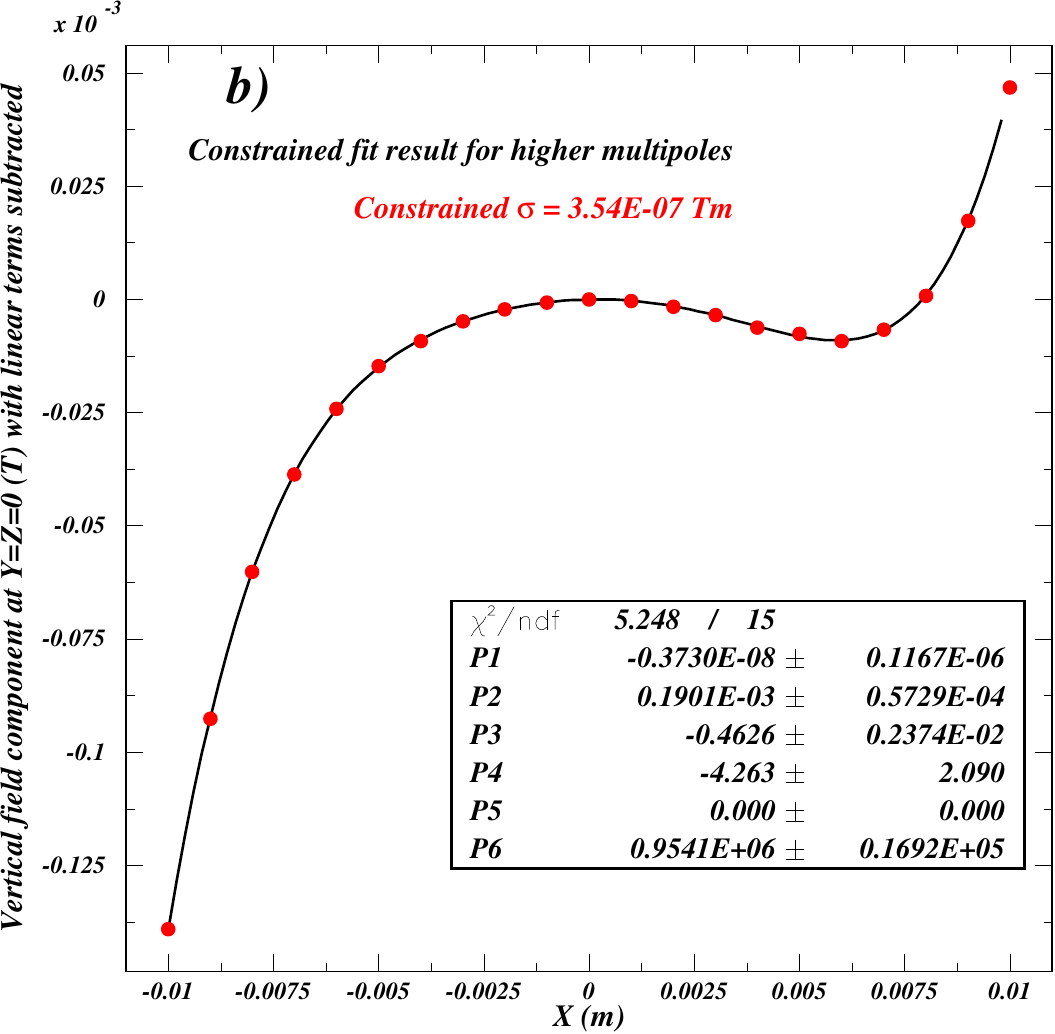}
        \caption{a)~Application of the procedure for determining higher multipole content to the south arc quadrupole Type~B shows that the extraction channel introduces sextupole and decapole terms in the integral of the vertical field component. b)~The result shown here for the multipole analysis to the central profile of the vertical field component indicates that the fringe field compensates (reduces) both the sextupole and dodecapole terms caused by the extraction channel at the center of the magnet by about a factor of four.}
        \label{fig:fittypeb}
    \end{figure}

Applying the above procedure to the field calculation for the quadrupole Type~B
can now identify the effect of the extraction channel on the higher multipole content. The result for the field integral shows
significant sextupole and decapole terms in addition to the octupole and dodecapole terms, as shown in Fig.~\ref{fig:fittypeb}~a).
The octupole and dodecapole terms are of magnitude similar to those of the Type~A quadrupole, though the octupole term is of opposite sign and much better determined.
    Figure~\ref{fig:fittypeb}~b)
    shows that the sextupole term in the integral is significant, but about a factor of four smaller than at the longitudinal center of the magnet. The same is true of the dodecapole term. The decapole term is due to the fringe field. The fringe field overcompensates the octupole term, flipping its sign. The small decapole term is introduced by the fringe field.

    To summarize this section on multipole analysis, we have used the example of the
    final, fully refined, model to exemplify the accuracy with which the multipoles in the field integral have been determined. Multipole coefficients are determined at the level of a few percent. The field value fluctuations around the polynomial fit are about $10^{-7}~$Tm, which is about $10^{-6}$ of the field integral at ${\rm X} = 0.01$~m. Section~\ref{sec:feamodel} will describe how the various steps in the refinement of the finite-element model lead to this result. We note that the required degree of refinement depends crucially on the magnet geometry. In our present case of a well-defined pole tip shape and a magnet length much greater than the bore radius, we require a highly accurate field calculation to determine the small higher multipole contributions.

\subsection{Numerical Violations of Maxwell's Equations}
\label{sec:diagnostics:maxwellviolations}
\begin{figure}[htb]
      \centering
\includegraphics[width=0.7\columnwidth]{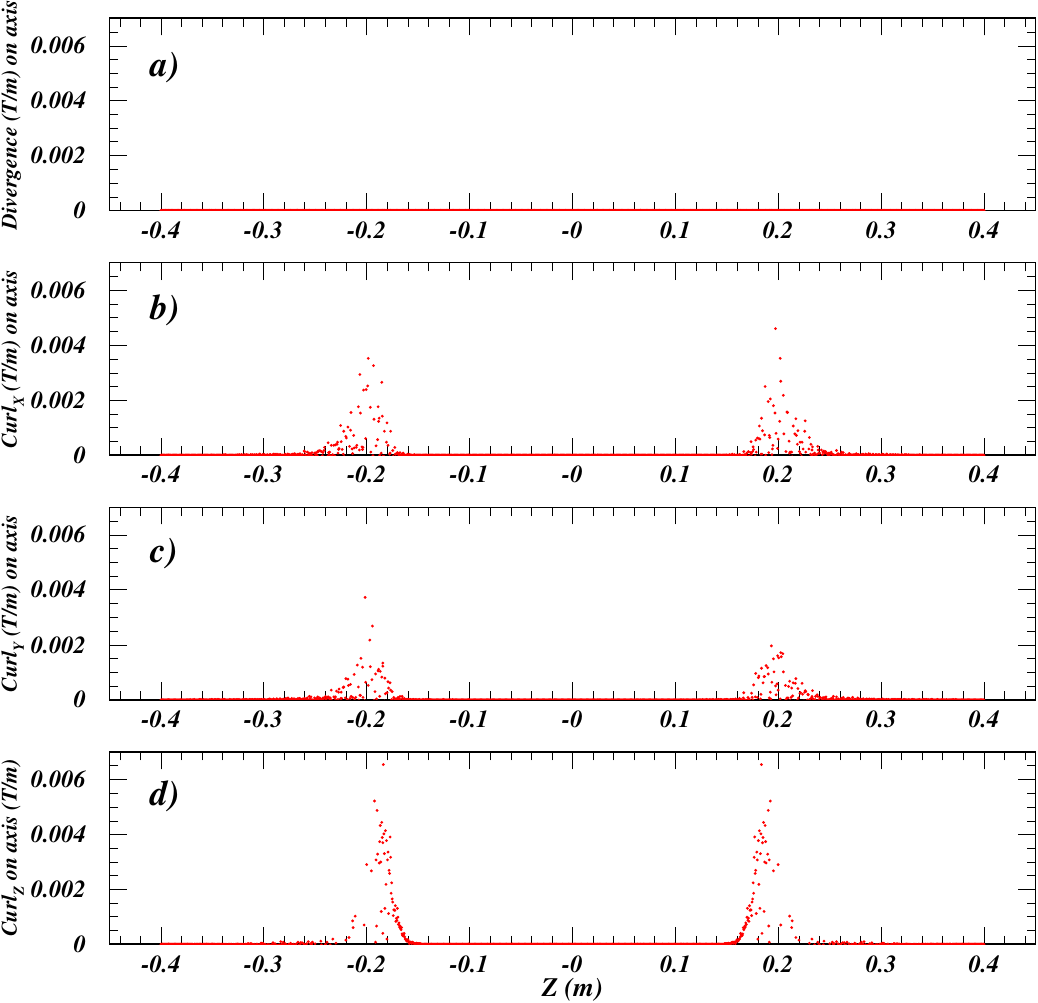}
        \caption{Values of a)~the divergence and the b)~X, c)~Y, and d)~Z curl components as functions of Z on the magnet axis for the quadrupole Type~A. While the divergence is accurate at machine accuracy, the curl components exhibit errors in the fringe field regions.}
        \label{fig:divcurl}
    \end{figure}

\begin{figure}[htb]
      \centering
      \includegraphics[width=0.7\columnwidth]{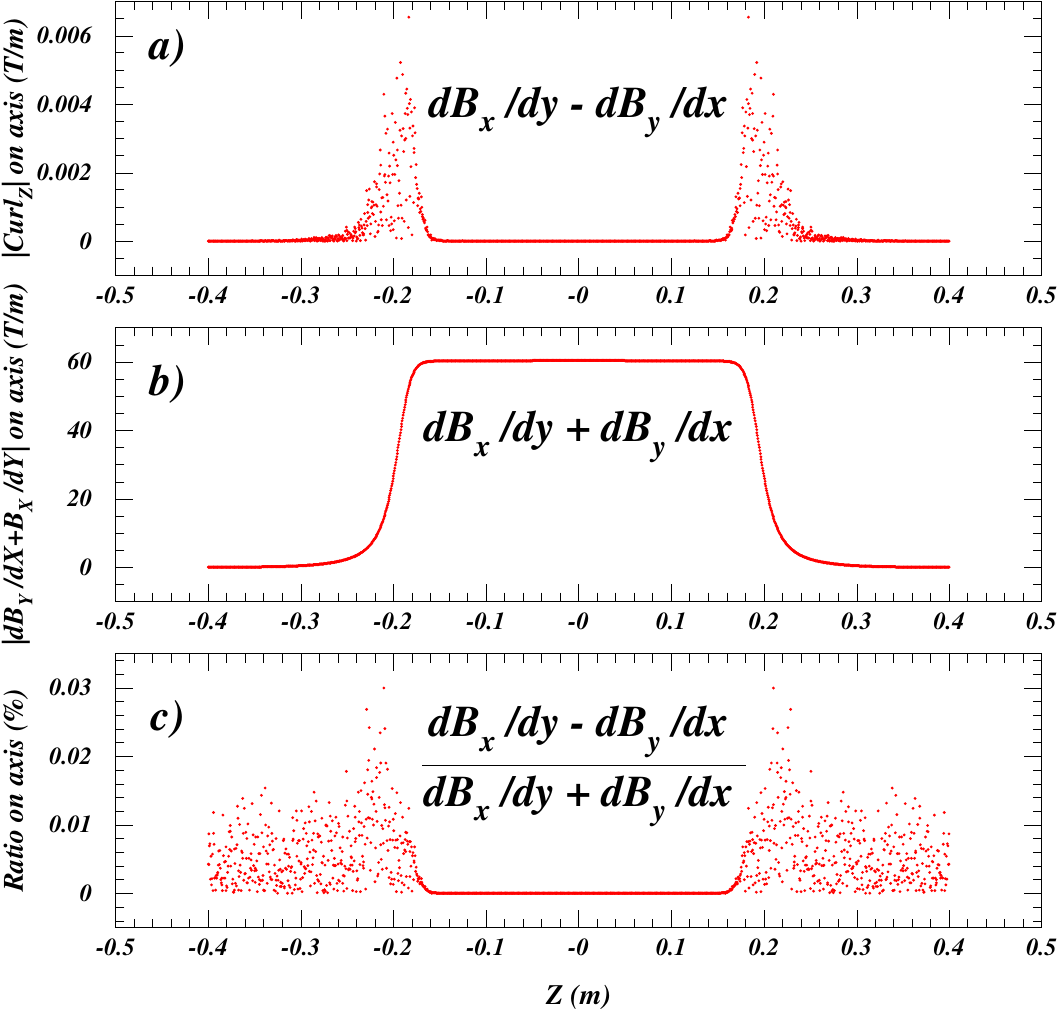}
      \caption{Absolute values of a)~the difference, b)~the sum and c)~the ratio of the vertical and horizontal field derivatives as functions of the longitudinal coordinate~Z at X=Y=0.}
        \label{fig:curlz}
    \end{figure}

The second means of quantifying numerical sources of error in the field calculation to be illustrated is the deviation from Maxwell's equations. Since there are no sources in the region of interest to particle tracking, the values of the divergence and curl should be small in an accurate finite-element model. The values for the derivatives of the field components are calculated for each point in the field map using the
two adjacent points in each of the three coordinates. Edge values are omitted in
the following plots. Again using the example of the final refined model, we show in Fig.~\ref{fig:divcurl}
    the values of the divergence and the curl components as a function of the longitudinal coordinate~Z on the magnet axis for the quadrupole Type~A. The divergence values are accurate at a level similar to machine accuracy, but the curl values show errors at the $5 \times 10^{-3}$~~T/m level in the fringe field regions. For any given application, any of these four error sources may be of interest, but for our present purposes of illustration, we concentrate on the Z~component of the curl.
Figure~\ref{fig:curlz}
shows the longitudinal dependence of the difference, sum and ratio of the vertical and horizontal gradients on the magnet axis. The difference is the Z~component of the curl, and the sum is close to twice the field gradient. The relative errors reach 0.03\% in the near exterior of the magnet. It should be noted that this is not a general characteristic of all models, or even of this one off axis. For example, the relative errors in the Z~component of the curl at $X=1$~cm in the body of the magnet are comparable to those outside of the magnet, reaching a level of about 1\%.

\begin{figure}[htb]
      \centering
      \includegraphics[width=0.6\columnwidth]{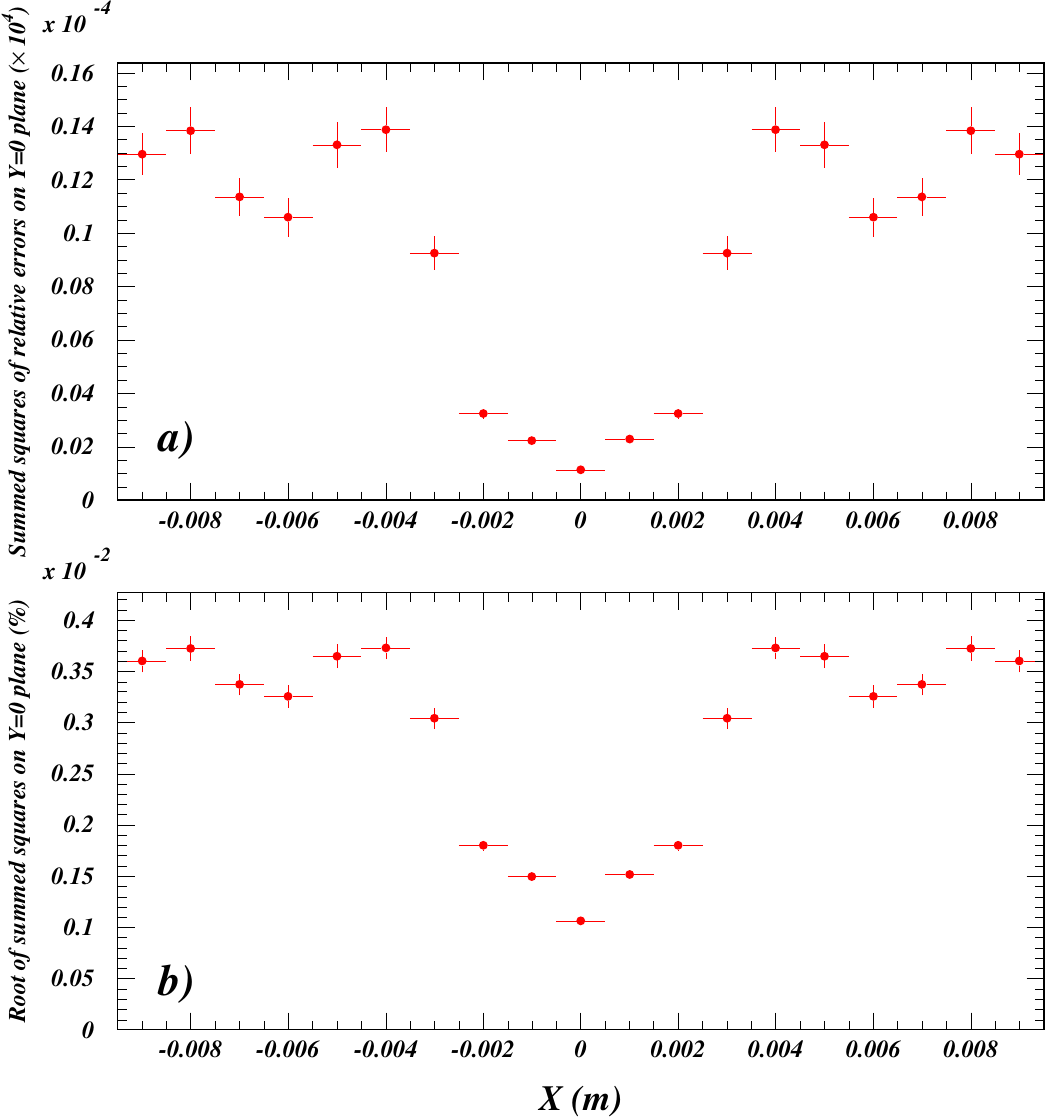}
        \caption{Results for a)~the sum over~Z of the squared ratios of the difference of $dB_{\rm Y}/d{\rm X}$ and $dB_{\rm X}/d{\rm Y}$ to their sum, and b)~the square root of the summed squares, as a function of the horizontal coordinate X.}
        \label{fig:sumsquares}
    \end{figure}

\clearpage
Having obtained a distribution in~Z of relative errors for all X~positions in the field map, we can plot this measure of field error as a function of~X. We choose to use the sum over Z of squared errors, since each contributes an error in particle tracking, regardless of sign.\footnote{Tracking results suffer in addition the accumulation of errors along the trajectory, which we do not account for here.}
For this reason, the simple sum would provide a misleading measure due to cancellations. Also, an uncertainty in the sum of squares of relative errors is readily obtained, providing a value for the uncertainty in the error obtained for each X~value. The results for the summed squares and their square root as  functions of~X are shown in Fig.~\ref{fig:sumsquares}.
    The vertical scale in~Fig.~\ref{fig:sumsquares}~a) is in units of $10^{-4}$ to aid the comparison to ~Fig.~\ref{fig:sumsquares}~b), where the vertical scale is in percent.

\begin{figure}[htb]
      \centering
      \includegraphics[width=0.7\columnwidth]{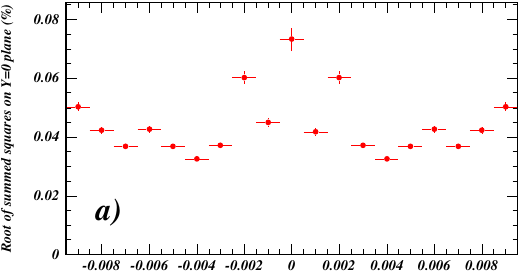}
        \linebreak
        \includegraphics[width=0.7\columnwidth]{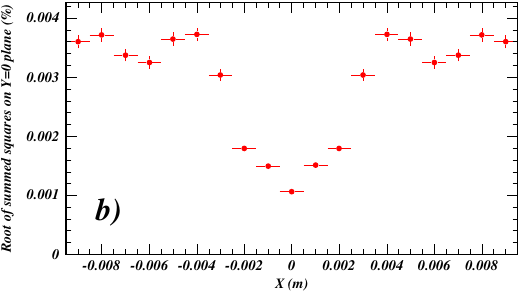}
        \caption{The square root of the summed squared ratios of the difference of $dB_{\rm Y}/d{\rm X}$ and $dB_{\rm X}/d{\rm Y}$ to its sum
          as a function of the horizontal coordinate~X for a)~the initial model, and b)~the final finite-element model version~6. See Sec.~\ref{sec:feamodel} for details on the successive refinements of the finite-element model. One can conclude that the field errors near the magnet axis have been reduced by  a factor of more than~70.}
        \label{fig:v6tov1}
    \end{figure}

    Thus we have obtained a quantitative measure of the field calculation errors which can be compared from model to model during the refinement of the finite-element mesh. Such a comparison
    is shown in Fig.~\ref{fig:v6tov1}, where the refined version~6 is compared to the initial version~1 (see Sec.~\ref{sec:feamodel} below for elaboration). 
    One can conclude that the refinement process emphasized the region of interest near the beam
    on the magnet axis and reduced the relative error by a factor of more than~70.

\begin{figure}[htb]
      \centering
      \includegraphics[width=0.7\columnwidth]{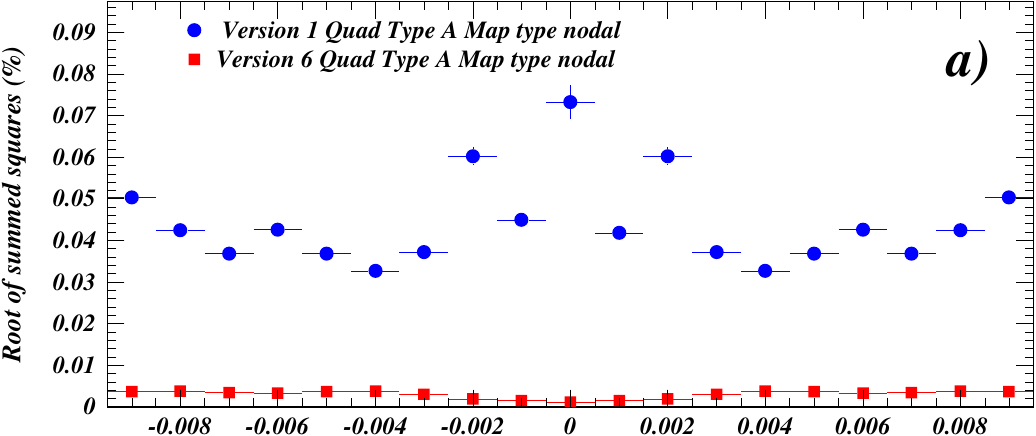}
      \linebreak
      \includegraphics[width=0.7\columnwidth]{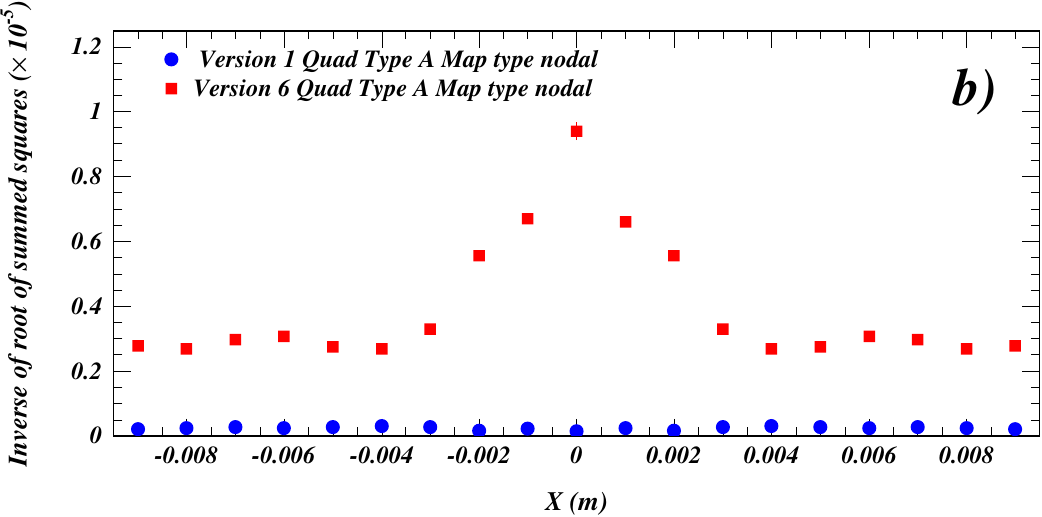}
        \caption{a)~Comparison of the results for the root of the summed squares of field errors for the two models shown in Fig.~\ref{fig:v6tov1}. This comparison has the distinct disadvantage of emphasizing the worse model over the better one. In order to clearly show the features of the more refined models, we will use the inverse of these values in our discussion of model development to follow below, as shown in~b).} 
        \label{fig:inverse}
    \end{figure}

    In order to compare the progress in multipole finite-element models of progressive refinement,
    we wish to show these results in a single figure, as shown in Fig.~\ref{fig:inverse}~a).
    However, this depiction requires a vertical scale which emphasizes the worst model, while we
    prefer the figure to clearly characterize the features of the improved model. For this reason, we choose to plot the inverse of the root of the summed squares of the relative errors, as shown in  Fig.~\ref{fig:inverse}~b), in our discussion of model development below in Sec.~\ref{sec:feamodel}.


   \clearpage
\section{Finite-element Model Development}
\label{sec:feamodel}
\subsection{Initial Model}
\label{sec:feamodel:initialmodel}
\begin{figure}[htb]
      \centering
      \includegraphics[width=0.7\columnwidth]{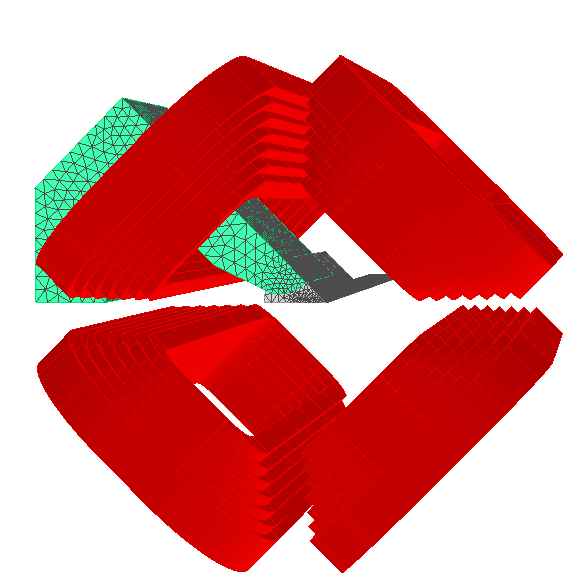}
        \caption{Geometry and finite-element mesh of the initial model (version~1). Note the 1/16~volumes for the air and steel, taking full advantage of the quadrupole symmetry and the symmetries
        about the XY, XZ, YZ and 45-degree planes.}
        \label{fig:nemesh}
    \end{figure}

We begin our review of the development of the finite-element model for the CESR south arc
quadrupoles with a description of the initial model, which we call Version~1. This model takes advantage of the full quadrupole symmetry of type A, so it is a 1/16 model as shown in Fig.~\ref{fig:nemesh}.
This geometry allows for tangential magnetic boundary conditions on the XY and 45-degree surfaces of the steel and air volumes, and for normal magnetic boundary conditions on the XZ and YZ surfaces.
Three air volumes are defined: 1) a small volume around the magnet axis, 2)~a buffer volume providing transition between the fine mesh of the beam volume and the other volumes, 3)~a background volume
with a coarse mesh (not shown in Fig.~\ref{fig:nemesh}) which defines  the distance from the inner volumes to the tangential boundary conditions on its outer surfaces. The buffer volume is defined as a cylinder
with the pole steel cut away. There are two steel volumes: 1)~the intersection of a cylinder with the pole tip to allow the definition of a finer mesh in the pole tip than in the rest of the steel, and 2)~the rest of the magnet steel. The dimensions, maximum cell size specification and cell geometry type for each volume of the full model are listed in Table~\ref{tab:fevolumes}.  Maximum cell size parameters can be defined both  in volumes and on surfaces. 
  \begin{table*}
  \centering
  \label{tab:fevolumes}
  \caption{Finite-element mesh parameters of the five volumes in the initial model for the symmetric Type~A quadrupole. All dimensions are given for the full model.}
          {\small
            \setlength{\tabcolsep}{3pt}
             \begin{tabular}{|l|c|c|c|c|c|c|c|c|c|}
    \hline
    Volume & Width & Height & Length & Radius & Cell geometry & \multicolumn{3}{c|}{Maximum cell size} \\
    & (m) & (m) & (m) & (m) & type & \multicolumn{3}{c|}{(m)} \\
    \cline{7-9}
    &&&&&& Volume & XY & XZ \\
    &&&&&&        & plane & plane \\
    \hline
    Beam & 0.05 & 0.005 & 0.60 & - & Quadratic & \multicolumn{3}{c|}{0.002} \\
    \hline
    Buffer & - & - & 0.60 & 0.05 & Linear & \multicolumn{3}{c|}{0.01} \\
    \hline
    Background & 2.8 & 2.8 & 3.60 & - & Linear & \multicolumn{3}{c|}{0.05} \\
    \hline
    Pole steel & - & - & 0.40 & - & Quadratic & {0.005} & {0.005} & {-} \\
    \hline
    Yoke & 0.47 & 0.47 & 0.40 & - & Linear & \multicolumn{3}{c|}{0.01} \\
    \hline
  \end{tabular}
   }
\end{table*}
\subsection{Refinement Steps}
\label{sec:feamodel:refinementsteps}
Intermediate results for five refinement steps of the finite-element model were obtained
in order to understand the effectiveness and performance cost of each type of refinement.
\begin{itemize}
\item{\underline{Version~2}}
The maximum mesh size on the XZ and 45-degree surfaces of the beam volume was reduced from 2.0~mm to 0.5~mm. Table~\ref{tab:multipoles} below in Sec.~\ref{sec:feamodel:results} shows that the
model increased in size by an order of magnitude. Decreasing  the maximum mesh size in the full beam volume would have been impractical. The accuracy of the field calculation improved remarkably, reducing the measure of deviations from Maxwell's equations by a factor of 20, as shown in Fig.~\ref{fig:modeldevel}. However, the accuracy in the determination of the multipole coefficients is still not
stable, as shown by their fluctuations during further refinements of the finite-element mesh.
\item{\underline{Version~3}}
Again modifying maximum cell sizes on surfaces only, the maximum cell size
on the pole tip surfaces was reduced to 2~mm~ from~5~mm. Quadratic finite-element cell shape definitions were
also introduced on these surfaces.
The small changes in the diagnostic criteria show that the mesh was already sufficient for this nodal method of field calculation. It was, however, of decisive importance for the integration method,
as discussed in Sec.~\ref{sec:feamodel:results}.
\item{\underline{Version~4}}
The maximum cell size in the beam volume was reduced from 2~mm to 1~mm. The  maximum cell size in the buffer volume was reduced to 5~mm from 10~mm and the cell type was changed from linear to quadratic. The dramatic reduction in the fluctuations of the field values around the polynomial fit for the multipoles ($\sigma_{\rm Residuals}$ is reduced by more than a factor of 20) shows that refinement of the
mesh in the volume between the beam volume and the pole tips is important for the overall accuracy of the model.
\item{\underline{Version~5}}
An additional micro-beam surface on the XZ plane was added, just 4~mm wide and 600~mm long, with
a cell size of only 0.2~mm. This reduced the curl values on the beam axis by about a factor of three. Interpretation of the results for the multipoles and $\sigma_{\rm Residuals}$ are complicated by
a fluctuation in the fit which caused the octupole term to be eliminated.
However, the stability in the determination of the dodecapole term is encouraging. It is determined with an accuracy of about 3\%.
\item{\underline{Version~6}}
The final refinement step was a test of the accuracy of the fringe field contribution.
A 100-mm-long, 50-mm wide XZ surface with a cell size of 0.4~mm longitudinally centered on the end of the magnet steel was introduced. The stability of the values for the integrated multipoles within their uncertainties indicates that their determination is not limited by the field calculation accuracy
in the fringe field region. 
\end{itemize}

\subsection{Results of the Refinement Steps}
\label{sec:feamodel:results}
\subsubsection{Multipole Analysis}
\label{sec:feamodel:results:multipoles}
The results of the multipole analysis described in Sec.~\ref{sec:diagnostics:multipoles} for
each of the six finite-element models are shown in the Table~\ref{tab:multipoles},
\begin{sidewaystable*}
  \small
  \centering
  \label{tab:multipoles}
  \caption{Results of the multipole analysis procedure applied to the field integrals
    $\int B_{\rm Y}({\rm X},{\rm Y}=0,{\rm Z})\;{\rm dZ}$ for the six finite-element models. The value for $\sigma_{\rm Residuals}$ is taken from the full polynomial fit after the insignificant terms have been constrained to zero. The uncertainties in these removed terms, used as a measure of the sensitivity
    of this procedure to their determination, are taken from the polynomial fit with the linear terms subtracted prior to removing the insignificant terms. The values given here for the non-zero multipole coefficients are obtained in the final step of the procedure, i.e. the constrained fit with the linear terms subtracted. Results are shown for the six nodal field calculation methods and also for the integration method in the fully refined model.}
  \setlength{\tabcolsep}{4pt}
  \begin{tabular}{|c|c|c|c|c|c|c|c|}
    \hline
    Version & Solver time & Model size & $\sigma_{\rm Residuals}$  & b$_2$ & b$_3$ & b$_4$ & b$_5$ \\
    & (minutes $\times$ cores) & (Nr of nodes) & (Tm) & (Tm/m$^2$)  & (Tm/m$^3$)  & (Tm/m$^4$)  & (Tm/m$^5$) \\
    \hline
    1n & 4.2 $\times$ 2 & 0.148M & $2.1 \times 10^{-5}$ & $(0) \pm 5.1 \times 10^{-1}$ &  $2.50 \pm 0.13 \times 10^3$ & $(0) \pm 5.2 \times 10^3$ & $-4.7 \pm 1.0 \times 10^{6}$ \\
    2n & 14.7 $\times$ 8 & 1.20M & $5.3 \times 10^{-6}$ & $(0) \pm 2.7 \times 10^{-3}$ &  $3.97 \pm 0.31 \times 10^2$ & $(0) \pm 4.14$ & $4.00 \pm 0.25 \times 10^{6}$ \\
    3n & 19.0 $\times$ 8 & 1.87M & $4.7 \times 10^{-6}$ & $(0) \pm 1.1 \times 10^{-1}$ &  $3.49 \pm 0.28 \times 10^2$ & $(0) \pm 1.1 \times 10^3$ & $3.46 \pm 0.22 \times 10^{6}$ \\
    4n & 36.9 $\times$ 12 & 2.50M & $1.9 \times 10^{-7}$ & $(0) \pm 2.5 \times 10^{-3}$ &  $-2.05 \pm 1.12 \times 10^2$ & $(0) \pm 2.6 \times 10^1$ & $3.0875\pm 0.091 \times 10^{5}$ \\
    5n & 31.0 $\times$ 24 & 2.88M & $9.3 \times 10^{-8}$ & $(0) \pm 2.2 \times 10^{-3}$ & $(0) \pm 5.3 \times 10^{-1}$ & $(0) \pm 2.2 \times 10^1$ & $2.861\pm 0.096 \times 10^{5}$ \\
    6n &  43.8 $\times$ 24 & 5.14M & $1.8 \times 10^{-7}$ & $(0) \pm 3.6 \times 10^{-3}$ &  $-1.26 \pm 1.07$ & $(0) \pm 3.7 \times 10^1$ & $2.963 \pm 0.086 \times 10^{5}$ \\
    6i &  43.8 $\times$ 24 & 5.14M & $4.1 \times 10^{-8}$ & $(0) \pm 1.0 \times 10^{-3}$ &  $-1.02 \pm 0.24$ & $(0) \pm 1.0 \times 10^1$ & $2.647 \pm 0.020 \times 10^{5}$ \\
    \hline
  \end{tabular}
\end{sidewaystable*}
together with the CPU time required for the TOSCA calculation and the number of finite-element nodes in the model. A measure of the scatter of the field calculation around the polynomial fit result, $\sigma_{Residuals}$ is taken from the full fit following elimination of the insignificant multipole terms (see Fig.~\ref{fig:fitprocedure}d) for the case of the fully refined model, version~6). The final results for the multipoles are taken from the last step of the procedure, i.e. the fit to the field values with the linear terms removed and the disallowed multipole terms constrained to zero. Since we are interested in the sensitivity to disallowed multipoles, we also include the uncertainty in their determination
from the initial fit with linear terms removed prior to setting them to zero (as shown in Fig.~\ref{fig:fitprocedure}c) for version~6). This is the fit result used to determine which terms to constrain to zero.
The accuracy obtained in the refinement process for the limit on the sextupole term b$_2$ (decapole term b$_4$) is $3.6 \times 10^{-3}\;\rm{Tm/m}^2$ ($3.7 \times 10^{1}\;\rm{Tm/m}^4$). In the convention sometimes used for the expression of multipole coefficients relative to the main quadrupole term, these values both correspond to 0.03 units at a radius of 0.01~m. The octupole (dodecapole) terms are found to be $-1.26 \pm 1.07\;\rm{Tm/m}^2$  ($2.963 \pm 0.086 \times 10^{5}\;\rm{Tm/m}^4$), or $-1.03 \pm 0.88 \times 10^{-1}$ ($2.423 \pm 0.0070$) units. The small octupole term carries a large relative uncertainty, but the dodecapole term is accurate
to~2.9\%.\footnote{We acknowledge that magnet fabrication errors at this level may be impractical and that their contribution to lattice errors may thus be larger. Our goal is the quantitative assessment of errors in finite-element models. We present a method to ensure that design model errors do not exceed the fabrication errors.}
\subsubsection{Deviations from Maxwell's Equations}
\label{sec:feamodel:results:maxwellviolations}
\begin{figure}[htb]
      \centering
      \includegraphics[width=0.7\columnwidth]{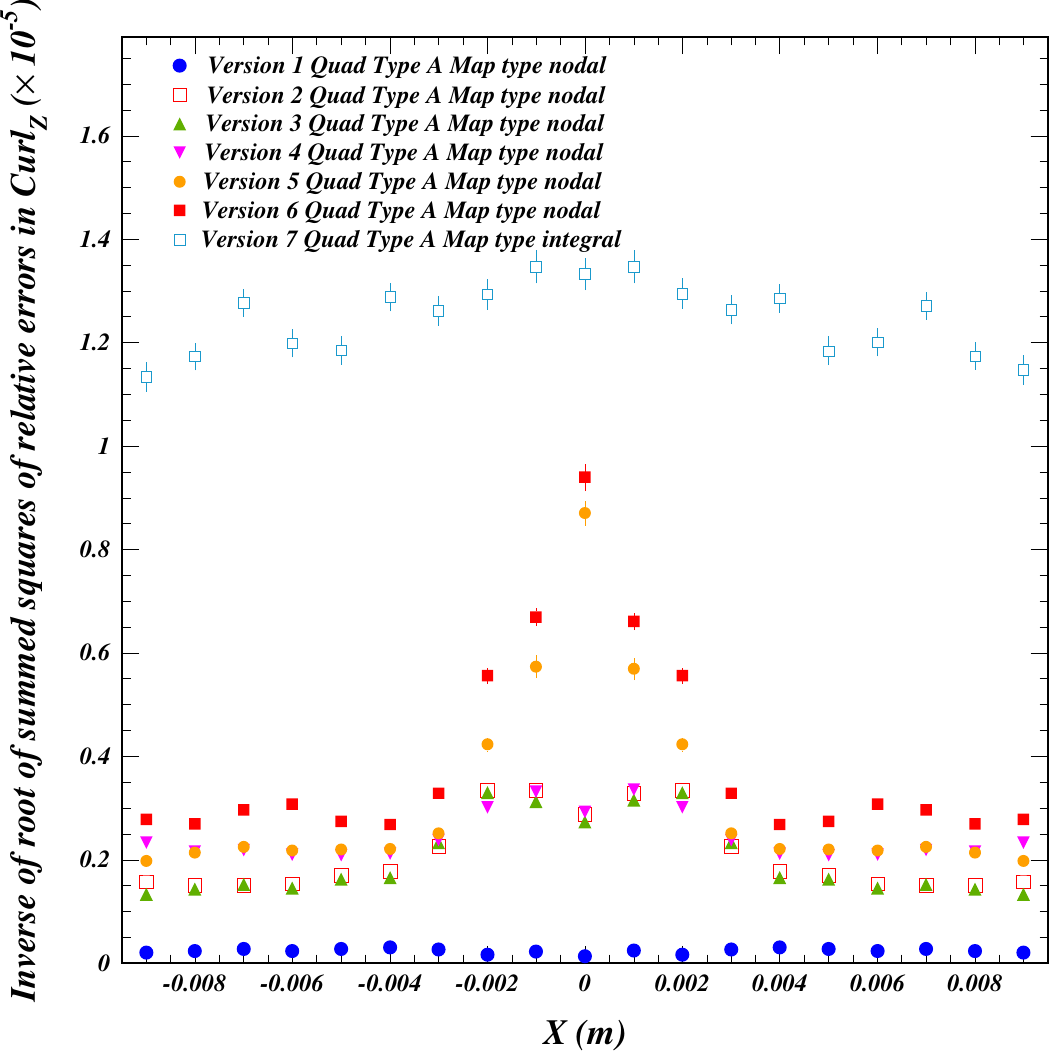}
        \caption{Improvement in the consistency of the field calculations with Maxwell's equations using the diagnostic criterion developed in Sec.~\ref{sec:diagnostics:maxwellviolations} for the six
          finite-element models described in Sec.~\ref{sec:feamodel:refinementsteps}.}
        \label{fig:modeldevel}
    \end{figure}

Figure~\ref{fig:modeldevel} 
shows the evolution of the diagnostic criterion developed in Sec.~\ref{sec:diagnostics:maxwellviolations} as a quantitative assessment of the level of numerical deviations from Maxwell's equations.
The relationship of these results to the particular mesh refinements is discussed above in Sec.~\ref{sec:feamodel:refinementsteps}. The refinement on the XZ and 45-degree planes within 2.5~mm of the magnet axis (version~2) resulted in a reduction  of the Z component of the curl by a factor of more than 20. The introduction of a ``micro-beam'' surface within 2~mm of the magnet axis of cell size~0.4~mm (version~5) significantly improved the accuracy in the region occupied by the positron beam, which is of the order of 1~mm wide wide in the south arc of CESR and is focused to a point near the end of each quadrupole. The final step (version~6) introduced a fine mesh volume in the fringe field region, significantly reducing the deviations from Maxwell's equations in the fringe region, but showing only a small effect on the field integrals.

The above analysis was performed for the results of the integration method as well as for the nodal method of field calculation and the results are included in Table~\ref{tab:multipoles} and in Fig.~\ref{fig:modeldevel} for the fully refined model. The post-processor computation time was more than 9~hours rather than the 13~minutes required by the nodal method. The decisive step in the mesh refined process for the integration method was in Version~3, when the mesh was refined on the pole tip surfaces. All other refinement steps were inconsequential for this calculation method. In the final model, however, this
method of calculation provides significantly smaller deviations from Maxwell's equations and an improvement in the determination of multipole coefficients by a factor of more than three.


\section{Conclusions and Outlook}
\label{sec:conclusions}
The methods developed during this work provide useful input to two types of
accelerator optics design, implementation, and operational diagnostic projects:
\begin{itemize}
\item
quantitative assessment of the accuracy of field maps derived from finite-element models
provides a diagnostic tool for the use of these field maps in particle tracking methods for
determining accelerator element transport matrices. These can prove useful at any stage of
accelerator development from calculating the effects of approximations and errors such as
misalignments during the design process to the diagnosis of deviations in measured optical functions from design.
\item
the reliable derivation of the multipole content in discrete field maps enables the implementation
of higher multipole coefficients in elements of lattice models for accelerator optics.
The orders-of-magnitude improvement in computing time makes possible many types of design and
diagnostic studies when field-map tracking is impractical. We have presented a
case of multipole determination for a quadrupole magnet design with very small multipole
content. The methods described here are generally applicable to finite-element models for
designs with large multipole terms arising from, for example, severe space constraints.
In such cases, both the tracking and multipole analysis will be of importance for both design and operational optics corrections. 
\end{itemize}

A prime example of the need for such studies are the splitter/combiner beam-lines for
the Cornell Brookhaven Energy-recovery-linac Test Accelerator\cite{cbetaprl2020,Gulliford:2020jxj}.
The compact nature of these beam-lines, together strict transverse uniformity specifications,
resulted in the use of 3D tracking models to design the transport magnets for the 42, 78, 114 and 150~MeV electron beams.
In fact, the design criteria were achieved through the introduction of angled
chamfers on the ends of the poles for the purpose of shaping the fringe fields, based on the
tracking results\cite{Crittenden:2018zgj,Crittenden:2017wur}. In addition to the two element model improvements
cited above, the Bmad library provides for fringe field shape parameters which were used effectively in the CBETA lattice
model\footnote{Private communication from J.S.~Berg of Brookhaven National Laboratory}.

An extensive study comparing various methods of producing and applying transfer maps,
including from discrete field tables, has been performed for the non-scaling fixed-field
alternating-gradient accelerator EMMA\cite{PhysRevSTAB.15.044001}.
Much work on avoiding numerical noise in generating reliable transfer maps from
discrete field maps by using Maxwell's equations to interpolate inward from a
bounding surface far from the magnet axis has also been
done\cite{Mitchell:ICAP2015-THBJI3,Mitchell:2011px,Mitchell:2010cu,Venturininima1999,mitchellthesis}.

We look forward in the near future to applying the model refinement and diagnostics
described here to both field-map tracking and implementation of multipole content
in the magnetic lattice elements for the relatively recently installed south arc
region of CESR. While we do not have specific expectations of surprises, we find
value in the results of such a first-time study, even when they
affirm the validity of past approximations for some applications.

\section{Acknowledgments}
We wish to acknowledge useful conversations with I.~Bazarov, G.~Hoffstaetter, V.~Khachatryan, D.~Rubin, D.~Sagan and S.~Wang, of CLASSE, as well as with J.S.~Berg, F.~M\'eot and N.~Tsoupas of Brookhaven National Laboratory. Valuable assistance was also provided by Y.~Zhilichev of Dassault Syst\`emes, including a critical reading of Sec.~\ref{sec:opera}.  J.~Shanks provided enlightenment concerning CHESSU optics design. Help was also provided by N.~Verboncoeur in the context of the National Science Foundation’s Research Experience for Undergraduates program, which is funded by award number PHY-1757811. This work is supported by the National Science Foundation award number DMR-1829070.

\raggedbottom


%

\end{document}